\documentclass[fleqn,twoside]{article}
\usepackage{espcrc2}

\usepackage{graphicx}
\usepackage[figuresright]{rotating}

\newcommand{\AmS}{{\protect\the\textfont2
  A\kern-.1667em\lower.5ex\hbox{M}\kern-.125emS}}
\def\lsim{\hbox{ \rlap{\raise 0.425ex\hbox{$<$}}\lower
    0.65ex\hbox{$\sim$}}}

\title{Gamma-Ray Bursts: Jets and Energetics}

\author{D. A. Frail\address[NRAO]{National Radio Astronomy
    Observatory, Socorro, NM 87801 USA}%
        \thanks{The NRAO is a facility of the
  National Science Foundation operated under cooperative agreement by
  Associated Universities, Inc.}}
  
\begin{document}

\begin{abstract}
  The relativistic outflows from gamma-ray bursts are now thought to
  be narrowly collimated into jets. After correcting for this jet
  geometry there is a remarkable constancy of both the energy radiated
  by the burst and the kinetic energy carried by the outflow.
  Gamma-ray bursts are still the most luminous explosions in the
  Universe, but they release energies that are comparable to
  supernovae. The diversity of cosmic explosions appears to be
  governed by the fraction of energy that is coupled to
  ultra-relativistic ejecta.
\end{abstract}

\maketitle

\section{Jet Signatures in Gamma-Ray Bursts}\label{sec:jets}

In hindsight, since jets are a natural outcome of most high energy
phenomena, they should have been expected in gamma-ray bursts (GRBs).
While there had been some early indications that relativistic outflows
from GRBs might not be isotropic \cite{wkf98,rho99b}, the real impetus
for invoking jets came from the ``energy crisis'' brought on by the
spectacular GRB\,990123 \cite{kdo+99}. On a timescale of order 80 s,
the isotropic energy released in gamma-rays E$_{iso}(\gamma)$ from
this burst approached the rest mass energy of a neutron star! If GRB
outflows were not isotropic but instead were collimated into jets with
an opening angle $\theta_j$ then they would only radiate into a
fraction $f_b =(1-\cos \theta_j) \cong \theta_j^2/2$ of the celestial
sphere \cite{rho99,sph99}. Thus the true gamma-ray energy released
$E_\gamma$ would be smaller than $E_{\rm iso}(\gamma)$ by the same
factor, i.e.~$E_\gamma=f_b\times E_{\rm iso}(\gamma)$.

\begin{figure}[t]
\centering
\includegraphics[scale=0.35]{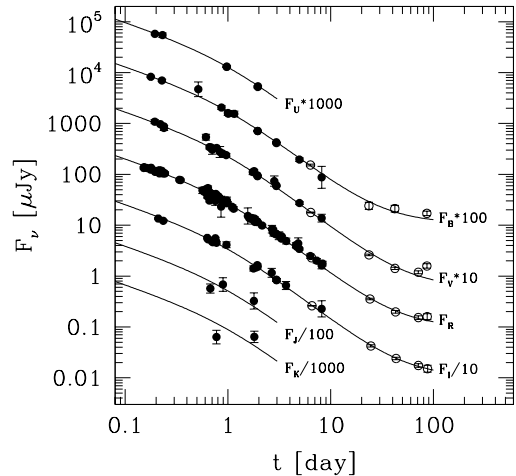}
\vspace{-35pt}
\caption{\it The UBVRIJK band lightcurves of GRB\,010222. The fits to the
  data are made with a broken temporal power law function added to a
  constant host galaxy component. The break in the light curves occurs
  at $t_j$=0.93 d. From \cite{grb+03}.
\label{fig:jet}}
\vspace{-25pt}
\end{figure}

The characteristic signature of a jet-like outflow at optical and
X-ray wavelengths is an achromatic break at a time $t_j$ in the
power-law decay of the light curves (Fig.~\ref{fig:jet}), in which the
exponent $\alpha$ (defined by $F_\nu\propto t^{\alpha}$) steepens by
$\Delta\alpha\sim{1}$ \cite{sph99}. At radio wavelengths, in which the
emission is initially produced by electrons radiating below the peak
of synchrotron spectrum $\nu_m$, a jet break is expected initially to
produce only a shallow power-law decay (e.g. t$^{-1/3}$ to t$^0$) of
the light curve until a time when $\nu_m$ passes through the radio
band. Thereafter, if the expansion remains relativistic, the power-law
decay index will be the same as the optical and X-ray
(i.e.~$\alpha\sim -2$). GRB afterglow measurements of the degree of
polarization and the polarization angle can provide a powerful
diagnostic for jets \cite{gl99,sar99}, but the observations to date
have been ambiguous.

The origin of this sharp break at $t_j$ is due to two effects. The
first is a purely geometric transition that occurs when the Lorentz
factor $\Gamma$ drops below $\theta_j^{-1}$, and the observer begins
to see the edge of the jet \cite{rho99}. A second effect that may
become important after $t_j$ is a dynamical transition in which the
jet outflow switches from a pure radial outflow to a laterally
spreading jet component. With some basic understanding of the jet
dynamics, a measurement of $t_j$ and redshift $z$ yields an estimate
of $\theta_j$ and hence $E_\gamma$. Readers who wish to learn more
about the practical difficulties in estimating $t_j$ and possible
systematic effects should review \cite{fks+01,bfs01,bfk03} and
references therein.

\section{A Standard Energy Reservoir for GRBs}\label{sec:standard}

In 2001 \cite{fks+01} we compiled all known bursts with $z$ and $t_j$
measurements (or limits) and derived their $\theta_j$ values using the
uniform jet model \cite{sph99}. The distribution of $t_j$ ranged from
1 to 25 d, while the derived $\theta_j$ values ranged from 2.5$^\circ$
to 17$^\circ$, with a mean of about 3.6$^\circ$. One immediate
implication, if GRBs are beamed to only a fraction of the sky, is that
the {\it true} GRB rate is a factor of $\langle f_b^{-1}\rangle\sim
500$ times the {\it observed} rate. Thus the prompt gamma-rays which
defines the GRB phenomenon is not observable by us for the vast
majority of bursts. For evidence of these off-axis explosions we must
look for ``orphan afterglows'' at optical and radio wavelengths
\cite{npg02,low+02,tp02}.

\begin{figure}[t]
  \centering 
\includegraphics[scale=0.37]{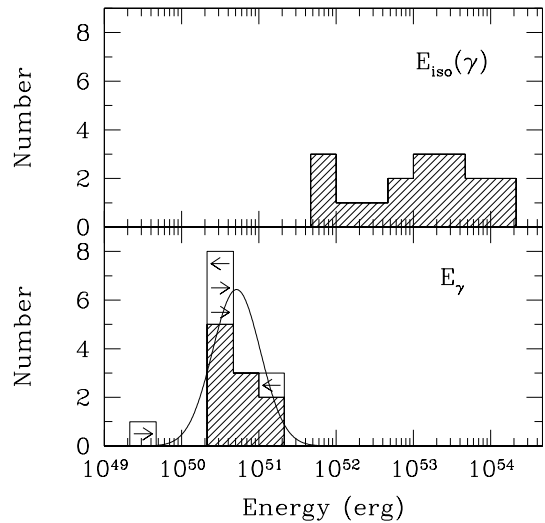} 
\vspace{-30pt}
\caption{{\it The distribution of the isotropic gamma-ray burst
    energy of GRBs with known redshifts (top) versus the
    geometry-corrected energy for those GRBs whose afterglows exhibit
    the signature of a non-isotropic outflow (bottom). The mean
    isotropic equivalent energy $\langle E_{iso}(\gamma)\rangle$ for
    17 GRBs is $1.1 \times 10^{53}$ erg, while the mean
    geometry-corrected energy $\langle {E_\gamma}\rangle$ is $5\times
    10^{50}$ erg.  Arrows indicate upper or lower limits. A
    circumburst density $n_\circ=0.1$ cm$^{-3}$ has been assumed. From
    \cite{fks+01}.}
\label{fig:fks}}
\vspace{-25pt}
\end{figure}

An even more remarkable result is that the isotropic gamma-ray energy
E$_{iso}(\gamma)$, which spans three orders of magnitude from
$\sim 10^{51}$ erg to $\sim{10^{54}}$ erg (Fig.~\ref{fig:fks}),
collapses into a narrow distribution of $E_\gamma$ once the geometric
corrections are applied. As shown in the bottom panel of
Fig.~\ref{fig:fks}, $E_\gamma$ clusters around $5\times 10^{50}$ erg,
with a 1-$\sigma$ multiplicative factor of only two. This constancy of
the product E$_{iso}(\gamma)\times \theta_j^2$ implies that the broad
distribution of gamma-ray luminosity (and fluence) which has been
observed is due in large part to the diversity in jet opening angles.

\begin{figure}[t]
\centering
\includegraphics[scale=0.45]{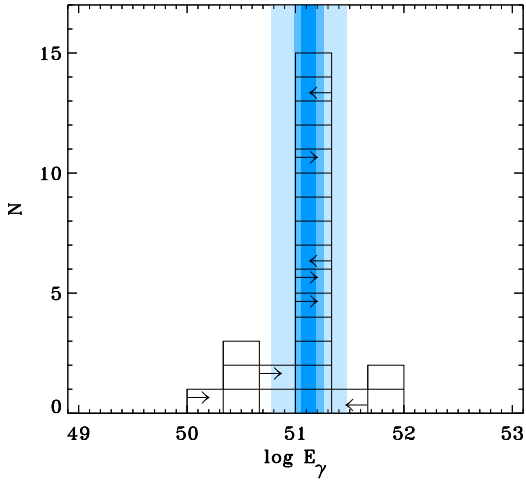}
\vspace{-40pt}
\caption{{\it The distribution of geometry-corrected energies with a  
    mean $\langle{E_\gamma}\rangle$=$1.3\times 10^{51}$ erg, a
    one-time sigma-clipped error of $\sigma = 0.07$\,dex.  Bands of 1,
    2, and 5 $\sigma$ about the standard energy are shown.  There are
    at least five identifiable outliers, more than 5$\sigma$ from the
    mean. From \cite{bfk03}.}
\label{fig:bfk03}} 
\vspace{-25pt}
\end{figure}

This clustering of $E_\gamma$ has been recently confirmed by an
improved analysis of a larger sample of bursts \cite{bfk03}. One
weakness of the earlier work \cite{fks+01} was that it adopted a
circumburst density $n_\circ=0.1$ cm$^{-3}$ for all bursts. With more
precise photometric data and the increasing sophistication of
afterglow modeling, the true range of $n_\circ$ is 0.1 cm$^{-3}$
\lsim\ $n_\circ$ \lsim\ 30 cm$^{-3}$ with a canonical value of 10
cm$^{-3}$ \cite{pk02}. The new result, shown in Fig.~\ref{fig:bfk03},
gives a mean geometry-corrected energy $\langle {E_\gamma}\rangle$ of
$1.33\times 10^{51}$ erg, with a 1-$\sigma$ multiplicative factor of
about two. The 2.7$\times$ increase in $\langle {E_\gamma}\rangle$
over the early value is almost entirely due to the use of realistic
$n_\circ$ estimates.

The sharply peaked $\langle {E_\gamma}\rangle$ distribution has
prompted suggestions that GRBs are ``standard candles'' and could
therefore have cosmographic applications. The scatter in $E_\gamma$ is
too small to place meaningful constraints on cosmological parameters
\cite{bfk03}. Moreover, the lack of local calibrators to pin down the
true $E_\gamma$ implies that the distribution of $E_\gamma$ at higher
redshifts only probes the shape of the cosmological Hubble diagram.
Energy diagrams like Fig.~\ref{fig:bfk03} are far more useful for
identifying potential sub-classes by their deviant energies.  Note
that $\sim$10-20\% of long-duration bursts are under-energetic in
gamma-ray energies. We will return to this point in
\S\ref{sec:diverse}.


\section{The Kinetic Energy of GRB Afterglows}\label{sec:kinetic}

The narrow $\langle {E_\gamma}\rangle$ distribution is puzzling, so it
would be useful to have independent check on this result. So far we
have used the gamma-ray energy as a proxy for the energy released by
the GRB explosion. However, only a fraction is radiated, while the
rest is carried away by the kinetic energy in the
outflow\footnotemark\footnotetext{The full GRB energy budget likely
  contains a significant amount of energy in neutrinos and
  gravitational waves, i.e.~$E_{tot}=E_\nu+E_{grav}+E_\gamma+E_k$.
  The early afterglow phase may also radiate away some additional
  energy.}. Particle acceleration occurring in this relativistic shock
gives rise to long-lived afterglow emission at X-ray, optical and
radio wavelengths \cite{mes02}. Thus with a suitable broadband model
that describes the dynamics of jet/circumburst interaction and
calculates the expected synchrotron and inverse Compton emission, all
the relevant quantities ($E_k$, $n_\circ$, and $\theta_j$) can be
calculated \cite{hys+01,pk01,fyb+03}.  Although high quality,
panchromatic datasets are rare and the validity of some of the
underlying model assumptions have yet to be fully tested
\cite{yost03}, the derived values of $E_k$ range from $10^{50}$ to
$3\times 10^{51}$ erg \cite{pk02}.

\begin{figure}[t]
\centering
\includegraphics[scale=0.4]{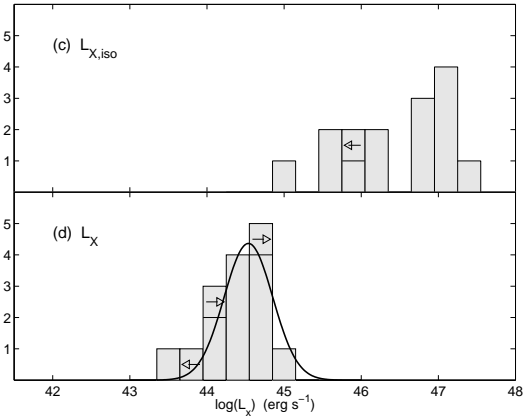}
\vspace{-25pt}
\caption{{\it The distribution of the isotropic X-ray luminosity
    measured at 10 hrs after the burst (top) versus the
    geometry-corrected X-ray luminosity (bottom). The narrowing of
    this distribution implies that the kinetic energy in the outflow
    is also approximately constant. From \cite{bkf03}.}
\label{fig:bkf03}}
\vspace{-20pt}
\end{figure}

A simpler approach is to exploit the fact that the flux density above
the synchrotron cooling frequency $\nu_c$ is proportional to the
energy per unit solid angle and the fraction of the shock energy
carried by electrons $\epsilon_e$ \cite{kumar00,fw01}.  The advantage
of this method is that it is insensitive to the density of the
circumburst medium or any other micro-physics in the shock, provided
that the emission is predominantly synchrotron. Thus X-ray afterglows,
which radiate above $\nu_c$ on timescales of several hours after the
burst, yield the product of $\epsilon_e\times E_k$, once $\theta_j$ is
known. This approach was recently carried out on a sample of X-ray
afterglows using $\theta_j$ values from \cite{bfk03}.  The results,
summarized in Fig.~\ref{fig:bkf03}, show the dramatic narrowing of the
X-ray luminosity $L_X$ as the geometric corrections are applied
\cite{bkf03}.  Although this method does not give $E_k$ directly, the
strong clustering of $L_X$ does imply that GRB explosions have a near
standard kinetic energy yield.

\begin{figure}
\centering
\includegraphics[scale=0.37]{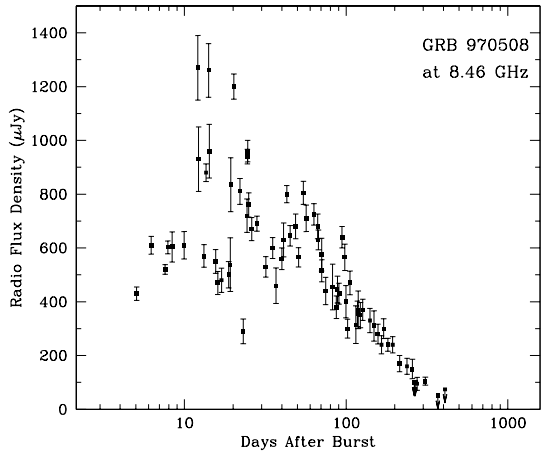}
\vspace{-35pt}
\caption{{\it Radio light curve of GRB\,970508. The rapid fluctuations
    seen at early times are due to interstellar scintillation. At late
    times ($t>$100 d) the behavior of the light curve is consistent
    with a quasi-spherical fireball expanding sub-relativistically.
    This allows true calorimetry of the explosion to be carried out,
    yielding $E_k\sim 5\times 10^{50}$ erg. From \cite{fwk00}.}
\label{fig:fwk00}} 
\vspace{-25pt}
\end{figure}

There is one method for estimating $E_k$ that does not require that we
know the geometry of the outflow. At sufficiently late times the
relativistic blast wave becomes sub-relativistic \cite{wkf98,wrm97}.
For kinetic energies of 10$^{51}$ erg and circumburst densities of 1
cm$^{-3}$ this occurs on a timescale of order 100 d
(Fig.~\ref{fig:fwk00}), and it can be recognized by a flattening of
the light curves compared to a jet in the relativistic regime
\cite{fmb+04}. After this time the dynamical evolution of the shock is
described by the Sedov-Taylor solutions rather than the relativistic
formulation and the outflow is expected to be quasi-spherical. This
method allows us to do true calorimetry of the explosion, permitting
not only the energy to be inferred but also the circumburst density,
the magnetic field and the size of the fireball.  The radius can be
checked for consistency with the equipartition radius and the
interstellar scintillation radius.  This method has been used for
GRB\,970508 \cite{fwk00} and for GRB\,980703 (Berger, {\it
  priv.~comm.}), yielding $E_k\sim 5\times 10^{50}$ erg, in agreement
with other estimates.

\section{Jet Geometry: Uniform vs Universal}\label{sec:universe}

The jet model adopted above assumed a uniform distribution of energy
(and Lorentz factor) per unit solid angle across the face of the jet,
which quickly drops to zero when the observer's viewing angle exceeds
the opening angle. Although this simple jet geometry provides a
straightforward way to estimate $\theta_j$ from $t_j$, it is unlikely
to be correct in practice. In collapsar simulatations of a
relativistic jet propagating through the stellar progenitor, the
Lorentz factor of the ejecta is high near the rotation axis, but
decreases off axis \cite{zwm03}. This has prompted an alternative
model in which GRBs have a structured jet configuration with the
energy per unit solid angle varying as
$\epsilon(\theta)\propto\theta^{-k}$, where $\theta$ is a viewing
angle that increases away from the jet symmetry axis. In this model
all jets have the same universal ``beam pattern'' and the breaks in
afterglow light curves at $t_j$ are a viewing angle effect.

Apart from its physical appeal, there are a number of distinct
advantages to the universal jet model. Well-known correlations between
gamma-ray luminosity, variability, spectral lag, and jet break time
\cite{nmb00,sg02,nor02} can be understood in the context of this
structured jet \cite{sal00,rl02}.  Moreover, the near-constant energy
result in \S\ref{sec:standard} can be preserved if there exists a
quasi-universal jet configuration for all GRBs with $k\simeq 2$
\cite{rlr02,zm02}. It is worth noting in this context that a prescient
paper \cite{lpp01} had earlier predicted both a quasi-universal jet
configuration and a standard energy yield for GRBs.

\begin{figure}[t]
\centering
\includegraphics[scale=0.5]{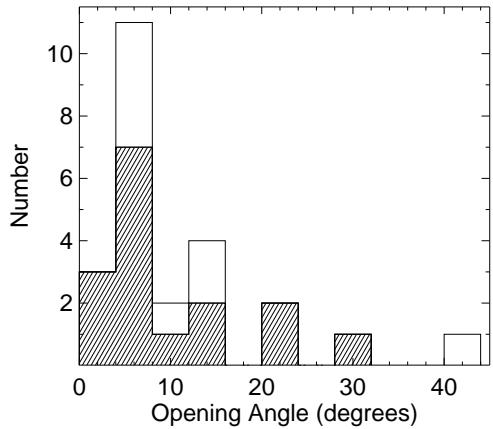}
\vspace{-35pt}
\caption{{\it The distribution of observed jet angles $\theta_j$ formed
    from data taken from \cite{bfk03}. Hatched squares represent true
    $\theta_j$ values, while unfilled squares are upper or lower
    limits on $\theta_j$. The structured jet model predicts a
    distribution of $\theta_j$ which is consistent with what is
    observed \cite{psf03}.
\label{fig:theta}}}
\vspace{-22pt}
\end{figure}

We have recently argued \cite{psf03} that a universal structured jet
can also predict the observed distribution of jet angles $\theta_j$.
One might naively expect that the number $dn(\theta)/d\theta$ of
bursts with angle in the interval $d\theta$, around $\theta$ would be
proportional to $\theta_j$, the observers viewing angle. This implies
that most bursts should have a large angle, contradicting the observed
distribution (Fig.~\ref{fig:theta}) which shows a peak near 6$^\circ$,
a deficit of narrow jets and a falloff in the number of wide-angle
jets, characterized by \cite{fks+01} as a power-law
$dn/d\theta\propto\theta_j^{-2.5}$. However, this argument ignores the
fact that bursts with small $\theta$ are brighter by a factor of
$\theta^{-2}$, and therefore can be seen (in a Euclidian universe) up
to a distance $\theta^{-1}$ farther, which contains a volume larger by
a factor of $\theta^{-3}$. Cosmological effects further limit the
volume at large redshifts, producing a cutoff in the number of the
narrowest jets. When these effects are accounted for, they reproduce
the observed distribution reasonably well and provide specific
predictions for future, more sensitive, gamma-ray experiments.

Despite the indirect evidence in favor of a universal structured jet
it remains far from proven. The resolution of this issue is an
important one since if affects estimates for the true GRB event rate
and the total energy, and it is crucial for possible unification
schemes of cosmic explosions (e.g. \cite{ldg03}). Fortunately, there
are a number of ways to discriminate between the uniform and
structured jet paradigms. As in the case of the opening angle
distribution $dn(\theta)/d\theta$, the structured jet model makes
specific predictions for the slope of the GRB luminosity function
which are best tested in future experiments \cite{rlr02,zm02}. Using
analytic and numerical hydrodynamic modeling, detailed predictions
have been made for Gaussian and power-law jet profiles (in both
$\Gamma$ and $\epsilon$) \cite{rlr02,sal03,gk03,kg03,pk03}. The light
curves produced by a jet in which {\it both} $\Gamma$ and $\epsilon$
vary as the inverse square of the angular distance from the jet
symmetry axis are almost indistinguishable from a uniform jet. With
other parameter values significant departures from the familiar broken
power-law are predicted, including variations in the sharpness of the
break with viewing angle and a flattening of the light curve before
and after the jet break time. Although some quantitative comparisons
have been made \cite{pk03}, we currently lack well-sampled photometry
around $t_j$ for a sufficient sample of bursts to properly constrain
jet structure. Polarization studies of afterglows show considerable
promise in this regard. Preliminary calculations \cite{rlsg02,sal03}
predict a peak in the degree of polarization near $t_j$ and (unlike
the uniform jet) predict no variation in the polarization angle with
time. The recent measurement of a time-variable position angle from
the polarized emission of GRB\,021004 has been used to argue against
the structured jet model \cite{rwf+03}. Since there was no
corresponding break in the light curves on the same timescale as the
position angle variations, we consider this issue unsettled.

\section{Cosmic Explosions: Quality vs Quantity}\label{sec:diverse}

As noted in \S\ref{sec:standard} there are true low energy outliers in
the distribution of $E_\gamma$ (Fig.~\ref{fig:bfk03}).  This includes
events like GRB\,980519 and GRB\,980326 which were classified as
fast-faders (f-GRBs) based on the steep decline of their optical light
curves ($F_\nu\propto t^{-2}$) at early times. The archetype of the
f-GRB sub-class is GRB\,030329 which exhibited a clear jet break at
optical (and X-ray) wavelengths at $t_j=0.55$ d \cite{pfk+03},
implying $E_\gamma\simeq 5\times 10^{49}$ erg - significantly below
the mean $\langle{E_\gamma}\rangle$=$1.3\times 10^{51}$ erg. The
peculiar GRB\,980425, associated with the Type Ic SN\,1998bw
\cite{gvv+98,kfw+98}, may define another possible sub-class of nearby
($d<100$ Mpc), low-luminosity events that are associated with bright
supernovae (S-GRBs) \cite{bkf+98,nor02}. Based on its gamma-ray
properties GRB\,980425 was severely under-energetic with
E$_{iso}(\gamma)\simeq 7\times 10^{47}$ erg. If we are to understand
this apparent diversity of cosmic explosions we must look more closely
at the {\it total} energetics of these peculiar events.

\begin{figure}[ht]
\centering
\includegraphics[scale=0.36]{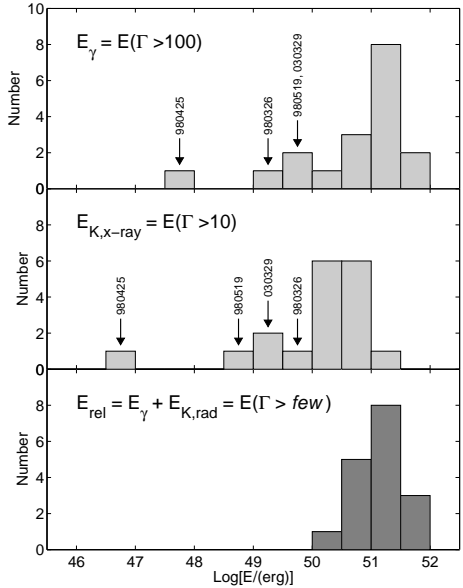}
\vspace{-22.5pt}
\caption{{\it Energy histograms. The top panel shows the distribution of 
    gamma-ray energy $E_\gamma$ and the center panel shows the kinetic
    energy $E_k$ inferred from X-rays (assuming $\epsilon_e=0.1$).
    Both of these methods are only sensitive to ejecta with large
    Lorentz factors ($\Gamma>10$).  Significant outliers are labeled,
    including GRB\,980425 and GRB\,030329. If the total relativistic
    energy $E_{rel}$ is derived instead (bottom panel), the dispersion
    narrows significantly as the energy outliers draw closer to the
    mean energy. This implies that cosmic explosions produce
    approximately the same quantity of energy, but the quality, as
    traced by ultra-relativistic ejecta, varies widely. From
    \cite{bkp+03}.
\label{fig:nature}}}
\vspace{-25pt}
\end{figure}

While the X-ray and optical observations of GRB\,030329 showed an
early jet break ($t_j=0.55$ d), the centimetre and millimetre light
curves \cite{bkp+03,sfw+03}, contrary to expectations, continued to
rise before exhibiting a break at ($t_j=9.8$ d). We proposed a two
component jet model with a narrow angle jet ($t_{NAJ}=0.55$ d and
$\theta_{NAJ}=0.09$ rad) which is responsible for the early afterglow,
and a wide angle jet ($t_{WAJ}=9.8$ d and $\theta_{WAJ}=0.3$ rad)
which carries the bulk of the energy in the outflow and dominates the
optical and radio emission after $\sim 1.5$ d \cite{bkp+03}. After
accounting for this lower Lorentz factor ejecta with a wider opening
angle, the total energy yield ($E_\gamma$ and $E_k$) is more in line
with estimates derived for ``typical'' bursts in \S\ref{sec:standard}
and \S\ref{sec:kinetic}. Similarly, for GRB\,980425, most of the
explosion energy does not appear to have been channeled into an
ultra-relativistic component. An analysis of the radio properties of
of SN\,1998bw gives convincing evidence for a relativistic component
with $\Gamma=2$-3, and a {\it minimum} energy $E_k\simeq 10^{50}$ erg
\cite{kfw+98,lc99}.

The hypothesis that emerges from this work is that cosmic explosions
(e.g. S-GRBs, f-GRBs, and GRBs) all draw from a standard energy
reservoir, but for reasons not currently understood, the fraction of
energy coupled to ultra-relativistic energy varies. Simply put, cosmic
explosions have the same {\it quantity} of energy, but the {\it
  quality} of that energy varies (Fig.~\ref{fig:nature}).

\section{Conclusions and Future Work}

For the last decade or more, progress in our understanding of GRB
progenitors has marched in lock step with the convergence of the GRB
energy scale.  By 1992 we knew from their peak flux and sky
distribution that GRBs were consistent with either a cosmological
population or a hitherto unseen population of sources in the halo of
our Galaxy. The resulting eight orders of uncertainty in the energy
scale led to a plethora of possible progenitor models \cite{nem94}.
With the demonstration that GRBs were cosmological \cite{mdk+97}, this
initial uncertainty shrank but the isotropic gamma-ray energy still
spanned three orders of magnitude -- reaching $\sim{10^{54}}$ erg --
creating a real strain on plausible burst models.

The solution, which seems obvious only in hindsight, was to recognize
that the relativistic outflows from GRBs are jet-like, not isotropic
(\S\ref{sec:jets}).  Consequently it appears that there is a near
standard energy yield of $\sim$10$^{51}$ erg in both the radiated and
the kinetic energy of the GRB explosion
(\S\ref{sec:standard}-\ref{sec:kinetic}). This result is all the more
remarkable when it is combined with strong evidence supporting the
view that long-duration GRBs are the result of the core collapse of a
massive star, aka ``collapsar'' \cite{smg+03,hsm+03}.  It is
convenient to think of a collapsar explosion as having two distinct
equal-energy components: quasi-isotropic ejecta expanding with
velocity $\Gamma\simeq 1.005$ producing the familiar radioactively
powered supernova light curves, and a highly collimated flow with
$\Gamma\simeq 100$ powered by synchrotron emission from electrons
accelerated in the relativistic shock. This picture, however, appears
to be too simple, since there is now growing evidence that the Lorentz
factor of the relativistic ejecta can vary as a function of viewing
angle and from one event to the next (\S\ref{sec:universe}). Energy
appears to be the one quantity that is (roughly) conserved throughout
(\S\ref{sec:diverse}).

By understanding the GRB energy budget we are gaining fundamental
insight into the inner workings of the central engine. As we move
forward studying new sub-classes of cosmic explosions, such as X-ray
flashes, events with on-going energy injection \cite{fyk+03}, and
bursts with distinct high-energy components \cite{gdk+03}, we would be
well-served to keep calorimetry as an essential tool.

\vspace{1pc}
{\bf Acknowledgements.} I would like to thank Ed van den Heuvel, who
made it possible for me to attend this meeting and give a final
farewell to {\it BeppoSAX} - The Little Satellite That Could. Much of
the work summarized in this paper was done by my collaborators,
including PhD students Edo Berger, Paul Price, and Sarah Yost.



\end{document}